\documentclass{ws-ijprai}
\usepackage{cite}
\usepackage{xurl}
\usepackage{hyperref}
\usepackage{graphicx}
\usepackage{csquotes}
\usepackage{amsmath,amssymb,amsfonts}
\usepackage{algorithmic}
\usepackage{textcomp}
\usepackage{framed,multirow}
\DeclareMathOperator*{\argmin}{arg\,min}
\usepackage{ mathrsfs }
\usepackage{float}
\usepackage{fixltx2e}

\begin{document}
\markboth{S. K. Sonbhadra \MakeLowercase{\textit{et al.}}}{ Learning target class feature subspace}

%
\catchline{}{}{}{}{}
%

\title{Pinball-OCSVM for early-stage COVID-19 diagnosis with limited posteroanterior chest X-ray images}

\author{Sanjay Kumar Sonbhadra}

\address{IIIT Allahabad, Prayagraj, U. P. India\\
\email{rsi2017502@iiita.ac.in}
}

\author{Sonali Agarwal}

\address{ IIIT Allahabad, Prayagraj, U. P. India\\
\email{sonali@iiita.ac.in}
}
\author{P. Nagabhushan}

\address{ IIIT Allahabad, Prayagraj, U. P. India\\
\email{pnagabhushan@iiita.ac.in}
}

\maketitle


\begin{abstract}
The infection of respiratory coronavirus disease 2019 (COVID-19) starts with the upper respiratory tract and as the virus grows, the infection can progress to lungs and develop pneumonia. The conventional way of COVID-19 diagnosis is reverse transcription polymerase chain reaction (RT-PCR), which is less sensitive during early stages; especially if the patient is asymptomatic, which may further cause more severe pneumonia. In this context, several deep learning models have been proposed to identify pulmonary infections using publicly available chest X-ray (CXR) image datasets for early diagnosis, better treatment and quick cure. In these datasets, presence of less number of COVID-19 positive samples compared to other classes (normal, pneumonia and Tuberculosis) raises the challenge for unbiased learning of deep learning models. All deep learning models opted class balancing techniques to solve this issue; which however should be avoided in any medical diagnosis process. Moreover, the deep learning models are also data hungry and need massive computation resources. Therefore for quicker diagnosis, this research proposes a novel pinball loss function based one-class support vector machine (PB-OCSVM), that can work in presence of limited COVID-19 positive CXR samples with objectives to maximize the learning efficiency and to minimize the false predictions. The performance of the proposed model is compared with conventional OCSVM and existing deep learning models, and the experimental results prove that the proposed model outperformed over state-of-the-art methods. To validate the robustness of the proposed model, experiments are also performed with noisy CXR images and UCI benchmark datasets.
\end{abstract}


\section{Introduction}

The respiratory coronavirus disease 2019 (COVID-19) is caused by severe acute respiratory syndrome coronavirus-2 (SARS-CoV-2) that is the most recently identified member of the coronavirus family \cite{sonbhadra2020target}. This deadly disease was initially reported during late December, 2020 and spread to all the countries worldwide \cite{stoecklin2020first}. World health organization (WHO) declared this infectious disease as a public health emergency of international concern (PHEIC) on January 30, 2020 as it reached to many countries \cite{SA1} and on Feb 11, 2020 named it \enquote{COVID-19}. On March 11, 2020 WHO declared this a pandemic \cite{world2020director}. After the fourteen months journey, this virus caused over 160.5 million infections and more than 3.3 million deaths worldwide till May 12, 2021 \cite{johnhop}. This fatal disease is highly infectious and to control its spread, following three ways have been suggested as most promising preventive measures: social distancing \cite{punn2020monitoring}, use of mask \cite{chowdary2020face} and early identification of infected people. The social distancing and mask are the established preventive measures that people are strictly following, but early diagnosis is still a challenge to the research community. COVID-19 is highly infectious, thus the active patient must be quarantined to break the chain of infection. The best known method of COVID-19 diagnosis is real-time reverse transcription-polymerase chain reaction (RT-PCR) that came into effect after the rapid antibody tests showed unreliable results, because antibodies appear after 9-28 days of the infection and by this time, an infected person may spread the disease, if not isolated \cite{cds}. The RT-PCR test is a costly and time taking process, whereas inability to early-stage diagnosis and due to rapid growth of infections, medical experts are continuously trying to search some other ways of diagnosis. 

SARS-CoV-2 virus initially affects the respiratory system of the infected person and in later stages, it affects lungs that may cause severe pneumonia \cite{damiani2020pathological}. Statistics show that  $\sim$14\% of the COVID-19 patients have shortness of breath and severe cough due to pneumonia, because as the viral infection increases, it damages the alveoli (small air sacs) and surrounding tissues. It is evident that the symptoms of COVID-19 are extremely heterogeneous, ranging from minimal symptoms to significant hypoxia with acute respiratory distress syndrome (ARDS) \cite{ma2020extracorporeal}. Like flu, the common early symptoms of COVID-19 are fever, dry cough, nausea, diarrhea, muscle aches, vomiting, headache, loss of smell or taste, sore throat, etc. COVID-19 starts affecting the lungs as soon as it reaches through the nose or throat. Statistics show that 40\%-45\% of COVID-19 patients remain asymptomatic, but the virus affects their respiratory system silently and resulting in severe pneumonia. 

Aiming to the diagnosis at initial stage; especially for asymptomatic patients, chest imaging may play a key role because the COVID-19 pneumonia is different from normal pneumonia and tuberculosis \cite{punn2020automated}. Although for lung-related disease diagnosis, chest CT imaging has been proven more effective, the CXR is preferred because comparatively it is widely available, faster and cheaper. Biomedical image segmentation and classification have become admired areas of research to make the present healthcare system more robust and responsive \cite{casiraghi2020explainable, agarwal2020unleashing}. In this context, the computing infrastructure advancements make it possible to apply the deep network approaches for complex biomedical image analysis tasks and especially, variants of convolution neural networks (CNN) are found very effective and efficient feature learning approaches in the field of biomedical image analysis \cite{tan2019efficientnet, krizhevsky2012imagenet}.

Recently, several deep‐learning based COVID‐19 detection techniques have been proposed \cite{punn2020automated} using collection of pneumonia CXR images from different datasets \cite{cohen2020covid}, \cite{irvin2019chexpert}, \cite{Oakden}, \cite{Stein}. The limited availability of COVID-19 positive samples may lead to biased outcome therefore, oversampling is opted as a solution by state-of-the-art deep learning approaches. The oversampling of medical images is not always a good practice to increase the classification accuracy, hence the present paper proposes a novel one-class classification (OCC) approach for COVID-19 diagnosis in the presence of a limited number of COVID-19 positive training samples. The identification of COVID-19 infection or pneumonia in CXR images is an anomaly detection task, which is specifically an one-class classification problem. It is evident that OCSVCs (support vector data description (SVDD) and OCSVM) are more suitable for anomaly/novelty detection tasks, where the negative class samples are totally absent \cite{alam2020one}. The OCSVCs are very sensitive to the noise, therefore the present research proposes an extended version of OCSVM \cite{scholkopf2001estimating} under the supervision of pinball loss function \cite{huang2013support}, named as PB-OCSVM. In present research work, CXR images of COVID-19 \cite{cohen2020covid}, radiological society of North America (RSNA) images \cite{Stein} and U.S. national library of medicine (USNLM) collected montgomery country - NLM(MC) \cite{jaeger2014two} datasets are utilized for experiments to justify the workability of proposed model. For experiments it is assumed that only COVID-19 infected CXR images are available for training, whereas other type of pneumonia and normal CXR images are totally absent at the time of training. The proposed PB-OCSVM possesses the following advantages:

\begin{itemize}
\item{} It is less sensitive to the noise compared to conventional OCSVM.
\item{} Its computational complexity is nearly equal to the conventional OCSVM, because pinball loss function does not add any extra computation overhead. 
\item{} PB-OCSVM reduces to the original OCSVM when pinball loss parameter tend to \enquote*{1}, and thus OCSVM can be considered as a special case of the proposed PB-OCSVM. 
\item{} PB-OCSVM computes a hyperplane similar to original OCSVM. 
\end{itemize}

The rest of the manuscript is organized as follows: Section \ref{l2} discusses the recent research contributions and Section \ref{l3} describes the utilized datasets. The proposed methodology is discussed in Section \ref{l4} whereas the experiments and results are discussed in Section \ref{l5}. Finally, the concluding remarks and future aspects are discussed in Section \ref{l6}.

\section{Related work}
\label{l2}

CXR images are always preferred for disease diagnosis due to wide availability and easy handling of X-ray machines. As discussed in the preceding section, the conventional way of testing COVID-19 via RT-PCR technique is a time taking and error prone process, especially for asymptomatic patients. Hence, the CXR image analysis is claimed to be a more robust alternate way to diagnose COVID-19. Concerning the same, several binary and multi-class classifications based deep learning models have been reported using collection of multiple CXR datasets \cite{cohen2020covid}, \cite{irvin2019chexpert}, \cite{Oakden}, \cite{Stein} with a common objective to develop a fully automatic diagnosis system, because the manual analysis is a time consuming process and require radiology experts. In this context, it is also evident with deep insight into literature that one-class classification approaches are not yet explored. The present research considers the diagnosis of COVID-19 infection as an anomaly or novelty detection task and proposes a novel variant of OCSVM named pin-ball OCSVM (PB-OCSVM), capable to work with limited number of training samples and robust against the noise. The proposed approach eliminates the need of oversampling of original samples as preferred by all deep learning approaches. This section gives comprehensive review of the most recently proposed deep learning approaches for COVID-19 diagnosis using CXR images.

\subsection{Deep learning approaches}

In the battle against the COVID-19 pandemic, researchers are continuously proposing artificial intelligence (AI) based  quicker and alternate approaches to identify COVID-19 infection from CXR images. It is evident with deep insight into recent COVID-19 related publications that deep learning models are capable to diagnose infections using CXR images, that supplant the conventional testing methods and helpful to reduce the growing burden on radiologists during this pandemic. It is also true that there is always a chance of infection while collecting samples for swab test, whereas CXR analysis is comparatively safer and easily manageable. 
Seeking to identify COVID-19 infection using CXR images, Narin et al. \cite{narin2020automatic} proposed deep CNN based ResNet50, InceptionV3 and Inception-ResNetV2 pre-trained transfer models. Later, Apostolopoulos et al. \cite{apostolopoulos2020covid} applied state-of-the-art CNN architectures based on transfer learning approaches for CXR image classification and for experiments, two different datasets have been used where images belong to three different categories: COVID-19, viral/bacterial pneumonia and normal cases. In the race of COVID-19 diagnosis, Khalifa et al. \cite{khalifa2020detection} offered generative adversarial network (GAN) based approach using CXR images while ensuring the robustness against the over-fitting by generating more images. The utilized dataset contains 5863 CXR samples of two classes: pneumonia and normal. For experiments, the GoogLeNet, Squeeznet, Resnet18 and AlexNet deep learning architectures have been experimented along with GAN, where Resnet18 along with GAN outperformed other models. Further, Sethy et al. \cite{sethy2020detection} introduced deep feature based support vector machine (SVM) model for COVID-19 infection classification of CXR images. It is found that SVM outperformed other models with extracted features from ResNet50. Later, effectiveness of ResNet-50 architecture was proved by Bukhari et al. \cite{bukhari2020diagnostic} under three categories of CXR images: non-COVID pneumonia, normal and COVID-19 pneumonia. The experimental results show that this approach efficiently diagnose COVID-19 infection. 

As an alternate way of COVID-19 infection detection, a customized ResNet-50 CNN based deep learning architecture named \enquote*{COVIDResNet} has been proposed \cite{farooq2020covid}. For experiments the input images are progressively resized to $128 \times 128 \times 3, 224 \times 224 \times 3$ and $229 \times 229 \times 3$ pixels and the network is fine-tuned at each stage with automatic learning rate selection, whereas the proposed model attained higher accuracy with minimum computational complexity. Later, Zhang et al. \cite{zhang2020covid} proposed a novel deep anomaly detection method for quicker and early diagnosis. To validate the performance of the proposed approach, CXR images of two different categories have been considered: COVID-19 positive cases and other pneumonia. To deal the class imbalance problem in the collected samples, authors proposed a CXR based COVID-19 screening model through anomaly detection task \cite{pang2019deep}. Most recently, Punn and Agarwal \cite{punn2020automated} proposed a transfer learning based deep learning architecture for COVID-19 diagnosis, whereas the class imbalance problem was handled via random oversampling and weighted class approaches have been used. The experimental results show that the proposed approach outperformed other state-of-the-art approaches.

Wang et al. \cite{wang2020covid} proposed a deep CNN for COVID-19 diagnosis called \enquote{COVID-Net} and experimented over COVIDx dataset. The proposed COVID-Net was pre-trained on the ImageNet dataset and then trained on the COVIDx dataset using the Adam optimizer. Performance (Multi-class classification) of this approach was compared with VGG19 and ResNet-50. Afterwards, Yujin et al. \cite{oh2020deep} proposed a patch-based ResNet-18 to work with minimum number of trainable parameters, where a novel probabilistic gradient-weighted
class activation map (Grad-CAM) method was also proposed that takes into account of patch-wise disease probability in generating global saliency map. The proposed method outperformed COVID-Net when the task was considered as multi-class classification problem. Later, Afshar et al. \cite{afshar2020covid} proposed a capsule network based approach for COVID-19 identification from CXR images and named it COVID-CAPS, where to deal class imbalance problem more weight is given to positive samples in the modified loss function. Experimental results show that the proposed method outperformed state-of-the-art approaches. In this context, Hall et al. \cite{hall2020finding} considered COVID-19 identification a binary classification task and  proposed an ensemble of the three types of CNN classifiers for COVID-19 diagnosis. In this approach pre-trained ResNet50 and VGG-16 models work with a newly proposed small CNN in presence of balanced dataset obtained by image augmentation.

From above discussion and deep insight into literature, it is evident that almost all the existing solutions for COVID-19 diagnosis reported till date follow the deep network architectures. Every deep learning model is data hungry and consumes massive computing resources as well as time. Hence, a faster diagnosis process is needed due to the rapid growth of COVID-19 cases and limited availability of healthcare experts (especially doctors and pathologists). Meanwhile, due to limited availability of COVID-19 CXR images, class imbalance appears the most critical issue that has been addressed by all recently proposed approaches with different solutions. Therefore, a more robust solution is desired, capable of working in the presence of limited number of positive class samples whereas the other class samples are absent. Naturally, the diagnosis of COVID-19 cases can be formulated as an anomaly or novelty detection task where only COVID-19 positive samples are available for training. With this notion, following are the key contributions of the present research:

\begin{itemize}
\item{} Formulated the COVID-19 diagnosis as a one-class classification problem.
\item{} Only COVID-19 positive CXR images are considered as training samples.
\item{} All training and other class samples are treated as test samples.
\item{} Performance comparison with recent deep learning approaches is done with following parameters: accuracy, precision, specificity, sensitivity and confidence\_interval.
\item{} Experiments are performed in noisy environment to show the robustness of the proposed model.
\item{} To show generalization performance of proposed model, 10 benchmark UCI datasets are also experimented.
\end{itemize}

\section{Dataset Description}
\label{l3}

\begin{figure}
	\centering
	\includegraphics[width=.8\linewidth]{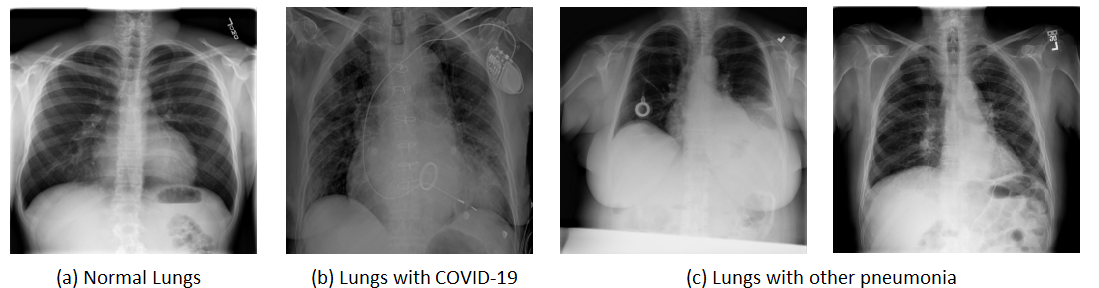}
	\caption{Sample CXR images.}
	\label{fig1}
\end{figure}

In present research, CXR images of COVID-19 \cite{cohen2020covid}, radiological society of North America (RSNA) \cite{Stein} and U.S. national library of
medicine (USNLM) collected Montgomery country - NLM(MC) \cite{jaeger2014two} datasets are collectively utilized for experiments to compare the performance of proposed method with existing state-of-the-art approaches. The COVID-19 dataset \cite{cohen2020covid} contains CXR images of following pneumonia classes: SARSr-CoV-1 (SARS), SARSr-CoV-2 (COVID-19), Pneumocystis spp., Streptococcus spp. and ARDS from multiple public domain resources available without infringing patient’s confidentiality (Fig.~\ref{fig1} (b)). The dataset includes the statistics up to August 2020 with metadata: patient ID, offset, sex, age, finding, survival, view, modality, etc. For experiments other samples from COVID-19 infected images are considered as another class named as \enquote*{other pneumonia}. The current version of the dataset is more rich in number of samples and classes, but to show the effectiveness of the proposed approach, the older version is used.

Another dataset used in this research has been introduced under RSNA pneumonia diagnose competition, which is a collection of 30,000 images chosen from the NIH CXR14 dataset \cite{Stein}. Among these samples, 15,000 images are pneumonia infected and out of remaining 15,000 samples, 7,500 had no findings whereas other 7,500 had different symptoms from pneumonia. Medical experts and radiologists annotated these images. Fig.~\ref{fig1}(c) shows sample instances. Initially, this dataset was published with 25,684 training and 1,000 test images and later, the 1000 test samples were merged to the existing training set to form 26,684 training images whereas 3,000 new images were introduced as test set. For more robust performance evaluation and comparative analysis, NLM(MC) \cite{jaeger2014two} dataset is also utilized that consists of 138 CXR images of classes: tuberculosis and normal. Table~\ref{tab1} shows the description of fused dataset used in present research. The dataset is composed of 1214 posteroanterior CXR images with classes: COVID-19 (108), other pneumonia (515), tuberculosis (58) and normal (533).

\begin{table}[]
\centering
\caption{Summary of utilized datasets.}
\label{tab1}
\begin{tabular}{|l|l|l|l|}
\hline
\textbf{Dataset}          & \textbf{Behaviour} & \textbf{CXR images} & \textbf{Total samples} \\ \hline
\multirow{2}{*}{COVID-19} & COVID-19             & 108                 & \multirow{2}{*}{153}  \\ \cline{2-3}
                          & Other pneumonia      & 45                  &                       \\ \hline
\multirow{2}{*}{RSNA}     & Normal               & 453                 & \multirow{2}{*}{923}  \\ \cline{2-3}
                          & Other pneumonia      & 457                 &                       \\ \hline
\multirow{2}{*}{NLM(MC)}  & Tuberculosis         & 58                  & \multirow{2}{*}{138}  \\ \cline{2-3}
                          & Normal               & 80                  &     \\ \hline
\end{tabular}
\end{table}

\section{Proposed approach}
\label{l4}

From preceding sections, it is inferred that deep learning approaches are playing a decisive role in advanced biomedical image analysis tasks. Several deep learning models have been proposed for COVID-19 diagnosis but suffer with following key issues:

\begin{itemize}
\item These models need massive computation power for training. 
\item Fails to work if samples of only one class (target class) are available for training. This phenomenon is the motivation of this research.
\item May give biased outcome if the dataset suffers with class imbalance problem.
\end{itemize}

\begin{figure}
    \centering
    \includegraphics[width=.8\linewidth]{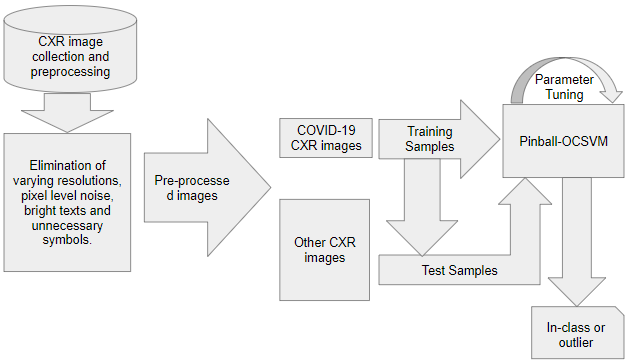}
    \caption{Schematic of proposed model}
    \label{fig002}
\end{figure}

The above issues are identified as the research gap and the present article proposes a one-class classification approach to solve these issues. The conventional OCSVM (a variant of OCSVCs) and the proposed PB-OCSVM are used as a one-class classifiers, capable of working with only target class (also known as class of interest (CoI)) samples. The OCC problem is different from conventional binary and multi-class classification tasks, where only target class is well defined and the other class samples are either totally absent or poorly sampled. In present research, only COVID-19 infected CXR images are used as training samples and the performance evaluation is done in two phases: considering all COVID-19 samples as test samples and considering other class images as test samples. The aim of the present research is to ensure maximum learning ability in presence of limited number of CXR images with minimum false-rejection and false-acceptance. Fig.~\ref{fig002} schematic of the proposed approach where it is assumed that only the target class samples are available for training. Initially, data pre-processing is performed to remove the noise from the images and then the processed images are fed to PB-OCSVM for training. After training, test samples (in-class and other class samples) are tested.

\subsection{Data pre-processing}
\label{pre}

In this article, COVID-19 infected posteroanterior CXR images of COVID-19 dataset \cite{cohen2020covid} are used to train the PB-OCSVM. For performance evaluation of the proposed model, the other pneumonia samples from COVID-19 dataset, RSNA \cite{Stein} and NLM(MC) \cite{jaeger2014two} datasets are used as test samples along with COVID-19 samples. The RSNA and NLM(MC) datasets contain samples of pneumonia and tuberculosis respectively along with normal cases. It is evident that the CXR images in the aggregated dataset suffers with undesired artifacts such as varying resolutions, pixel level noise, bright texts and unnecessary symbols, hence pre-processing is a necessary step for further analysis. To overcome from the textual and symbolic noise, in-painting is performed with the image mask generated using binary thresholding \cite{Openpunn} (Eq.~\ref{eq01}), whereas image resizing is performed for a fixed size resolution of $331 \times 331 \times 3$ (where 3 is number of channels) to work upon state-of-the-art models.

\begin{equation}
          N(x,y)=
     \begin{cases}
    {Th}_{max},& \text{i(x,y)} \geq  {Th}_{min}\\
    0,         & \text{otherwise}
  \end{cases}
   \label{eq01}
\end{equation}
where for input image $i(x,y)$, the mask is designed with minimum and maximum threshold values defined as ${Th}_{min}$ and ${Th}_{max}$ respectively. Even after filtering out undesired information, there is always a possibility of uncertainty at the deep pixel level representation \cite{hasinoff2014computer}. 

To preserve the original pixel value distribution, following two techniques are experimented to generate final pre-processed images: the adaptive total variation method \cite{szegedy2015going} and automatic color equalization (ACE) approach \cite{rizzi2003new}. Images of size  $331 \times 331 \times 3$ are used by these methods to generate final images.

\subsubsection{Adaptive total variation method}
Let, $\textit{I}$ is a grayscale image and $\gamma$ is a bounded set over ${\mathbb{R}}^2$, such that $\gamma \subset {\mathbb{R}}^2 $, denoising image $\mathcal{D}$ that closely matches to observed image $x=(x_1, x_2)$ $\epsilon$ $\gamma$ - pixels, given as:

\begin{equation}
\mathcal{D}  = \argmin_{\mathcal{D} }{\left ( \int_{\gamma}{}(\mathcal{D}  - I. \ln{\mathcal{D}})dx + \int_{\gamma}{}(\omega(x)|\Delta \mathcal{D}|dx) 
 \right)}
\label{eq4}
\end{equation}\\
where $\omega(x) = \frac{1}{1+ p{\mod{G_{\sigma}*\Delta \mathcal{D}}}'}$, $G_{\sigma}$ is the Gaussian kernel, $\sigma$ is variance , $p > 0$ is contrast parameter and * is convolution operator. 

\subsubsection{Automatic color equalization}

This method is based on a computational model of the human visual
system that merges the two basic global equalization mechanisms: \enquote{Gray World} (GW) and \enquote{White Patch} (WP). ACE is able to adapt widely varying lighting conditions and to extract visual information from the environment efficaciously. This is a two stage process where initially, the GW and WP approaches are merged via a type of lateral inhibition
mechanism, weighted by pixel distance that results in local-global filtering. Whereas the second stage maximizes the image dynamic via normalizing the white at a global level. 

Let the input image is $I$ with $c$ channels and initially, the chromatic/spatial adjustment produces as output image $R$ in first stage, where every pixel is recomputed according to the image content. Each pixel $p$ of the output image $R$ is computed separately for each channel $c$ as follows:

\begin{equation}
    \centering
    R_{c}(p) = \sum_{j \neq p} \frac{r(I_c(p)-I_c(j))}{d(p, j)}
\end{equation}
where $I_c(p)-I_c(j)$ decides the lateral inhibition mechanism, the distance $d(.)$ weights the amount of local or global contribution and $r(.)$ is the relative lightness appearance of the pixel. Like human visual system, the distance $d(.)$ weights the global and local filtering effect. Euclidean distance is considered for distance calculation. Relative pixel influence of each pixel is controlled by $r(.)$ and $d(.)$ for the spatial channel lightness adjustment. The variation of the slope of the function $r(.)$ acts as a contrast tuner and in the present research Signum function is used. Finally, the second stage linearly scales the range of values in $R_c$ independently in the relative channel $c$ into the range [0; 255] using the formula:
\begin{equation}
    \centering
    O_{c}(p) = round[s_c R_c(p)- m_c]
\end{equation}
for each pixel $p$ where $s_c$ is the slope of the segment
$[(m_c, 0); (M_c, 255)]$, with $M_c = max_{p} R_{c} (p)$ and $m_c = min_p R_c(p)$. To make this process more robust, dynamic of
the final image is always centered around the medium gray via means of White Patch/Gray World  (WP/GW) scaling as follows: 
\begin{equation}
    \centering
    O_{c}(p) = round[127.5 + s_c R_c(p)]
\end{equation}
For experiments, Euclidean distance and WP/GW scaling are used and $s_c$ is set to 20 for saturation function.

\begin{figure}[b]
\centering
\includegraphics[width=.8\linewidth]{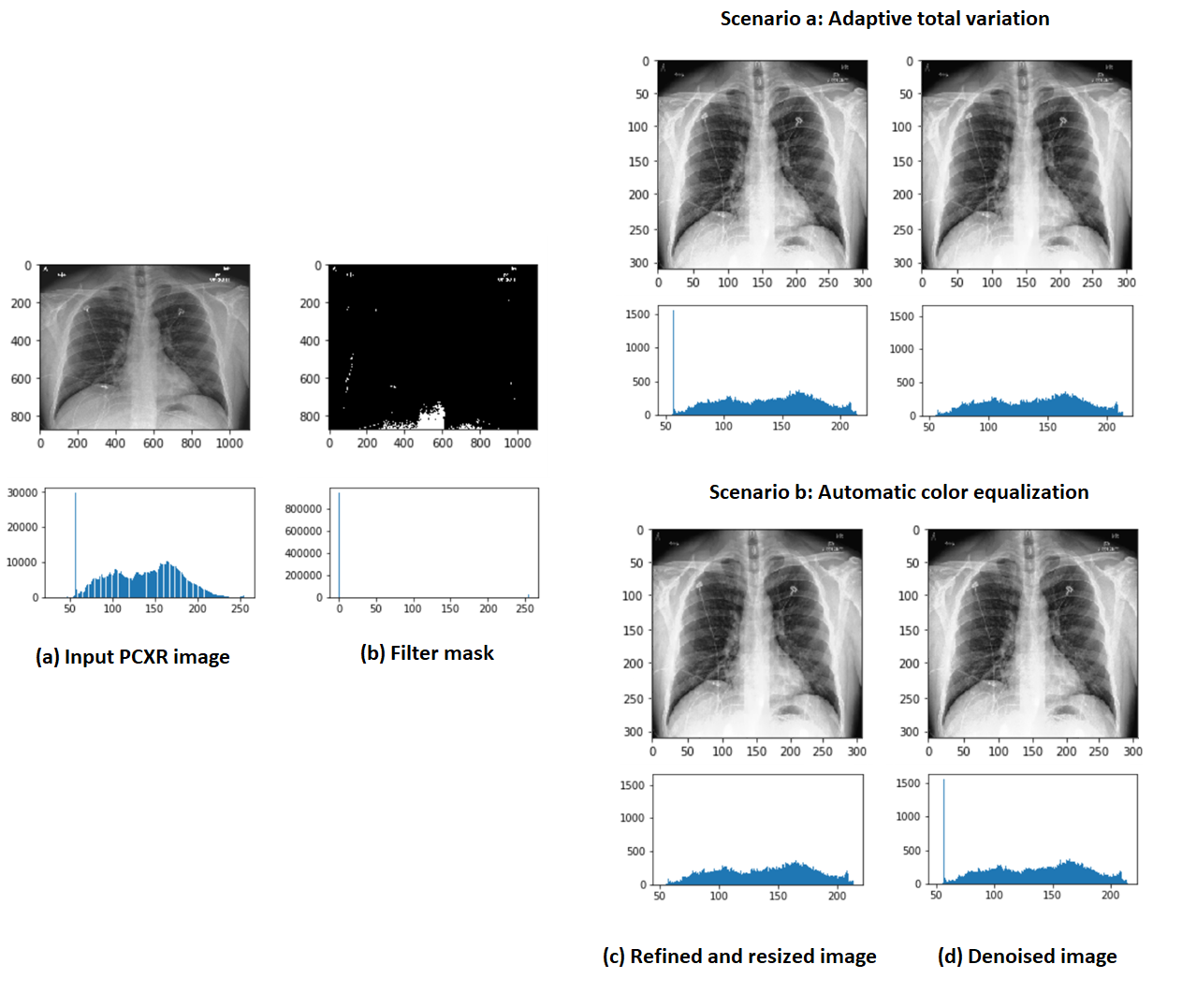}
\caption{Data pre-processing phases of chest X-ray image.}
\label{fig3}
\end{figure}

The complete data pre-processing operation is illustrated in Fig.~\ref{fig3}, where the histogram of resulting distributed pixels during every step of pre-processing is also shown. It is evident that the opted pre-processing approach is capable to eliminate the irregular intensities while preserving the original nature of the pixel distribution. After pre-processing of CXR images, intensive experiments are performed with PB-OCSVM, OCSVM and state-of-the-art deep network models. For PB-OCSVM and OCSVM, the COVID-19 positive samples are considered as training samples whereas deep learning models are trained with multi-class samples (80\% from each class) obtained via respective class balancing techniques. For OCC approaches, all in-class and other class samples are treated as test samples and for deep learning models all training, remaining in-class and other class samples are treated as test samples.

\subsection{Preliminaries}

This section gives brief description of OCSVM, variants of OCSVM with different loss functions and Pinball-SVM. This discussion will help to understand the formulation of proposed PB-OCSVM.

\subsubsection{OCSVM}

OCC algorithms are specifically used for anomaly/novelty detection tasks; especially when the non-target class is either ill-defined or totally absent. For unbiased operation of binary or multi-class classifiers, presence of two well-defined classes is necessary; but if the test sample belongs to unknown class, the classifier may exhibit biased behaviour. In such scenarios, OCC techniques are proven as robust solutions, majorly applicable for concept learning and outliers/novelty detection~\cite{khan2009survey}. 

\begin{figure}
	\centering
	\includegraphics[scale=0.3] {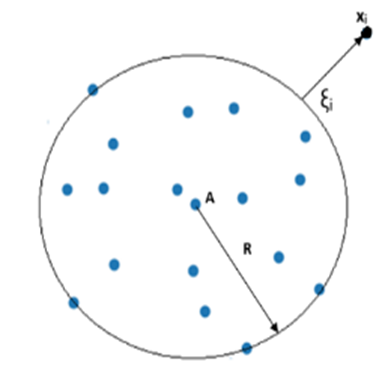}
	\caption{Support vector data description (SVDD).}
	\label{fig11}
\end{figure}

Tax et al.~\cite{tax1999support} proposed a novel OCC model termed as SVDD, where the target class samples are enclosed by a hypersphere, and the boundary points are called support vectors. If a data sample falls outside of the hypersphere, SVDD treats it as outlier and rejects (Fig.~\ref{fig11}). The objective function of SVDD is defined as:

\begin{equation}
\begin{aligned}
L(R,a,\alpha_i,\gamma_i,\xi_i)  = & R^2 + C \sum_{i} \xi_i -\sum_{i} \alpha_i \{R^2 + \xi_i- \\& (\parallel x_i\parallel^2 - 2a.x_i+ \parallel a\parallel^2)\} - \sum_{i}\gamma_i \xi_i 
\end{aligned}
\label{eq1000}
\end{equation}
subject \hspace{1mm}to: \hspace{1mm}$\parallel x_i-a\parallel^2\leq R^2+\xi_i,where\hspace{2mm} \xi_i\geq 0 \hspace{2mm} \forall$ i\\
where radius of the hypersphere is $R$, $x_i$  and $a$ are outlier and center of hypersphere respectively, the parameter \textit{C} controls the trade-off between the erros and volume, and $\xi$ is slack variable to penalizes the outliers. With the Lagrange multipliers $\alpha_i \geq 0$, $\gamma_i \geq 0$ where $i \in \{1, 2, \dots, N\}$, the objective is to minimize $R$ of the hypersphere while covering all target class samples with some penalty for outliers. By putting partial derivatives to zero and substituting the constraints into Eq.~\ref{eq1000}, following is obtained:

\begin{equation}
\centering
L= \sum_{i}\alpha_i(x_i,x_i) - \sum_{i,j} \alpha_i \alpha_j (x_i, x_j)
\end{equation}

If the description value of a test sample $x_i$ is greater than $C$ then it is treated as an outlier. Kernels can be used to reformulate the SVDD as follows:

\begin{equation}
\centering
{\left \|  { \phi (x) - a} \right \| }^2 \leq R^2
\end{equation}
The SVDD outcome can be computed as follows:
\begin{equation}
\centering
\label{p}
R^2 -{\left \|  { \phi (x) - a} \right \| }^2
\end{equation}
The output of Eq.~\ref{p} is positive for samples inside the boundary and negative for an outliers.

\begin{figure}
	\centering
	\includegraphics[scale=0.3] {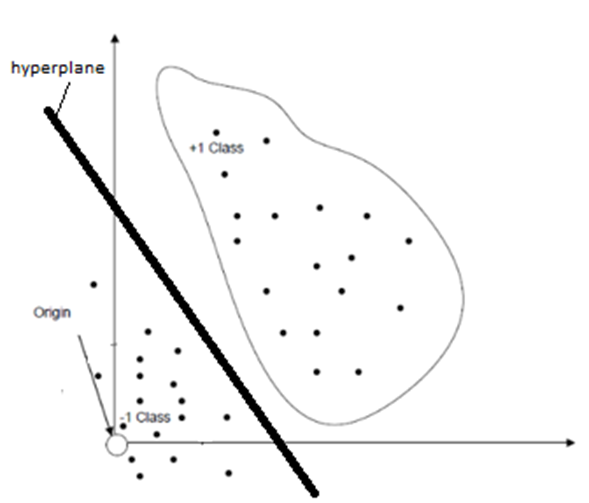}
	\caption{Working of OCSVM.}
	\label{fig22}
\end{figure}

Later, Schl{\"o}kopf et al.~\cite{scholkopf2001estimating} proposed one-class support vector machine (OCSVM) as an alternate OCC approach to SVDD, where a hyperplane separates the target class samples from outliers as shown in Fig.~\ref{fig22}. In OCSVM, class of interest (CoI) samples are separated by a hyperplane with the maximal margin from the origin, whereas the negative class samples reside in the subspace of the origin. The OCSVM is defined as following quadratic equation (Eq.~\ref{eq2000}):
\begin{equation}
\label{eq2000}
\max_{w,\xi , \rho}  \frac{1}{2} \parallel w \parallel^2+ \frac{1}{\upsilon N} \sum_{i}^{N} {\xi_i - \rho }
\end{equation}
subject to: $\hspace{2mm}  w. \phi(x_i )\geq \rho-\xi_i \hspace{2mm}$ and $\hspace{2mm} \xi_i\geq 0 \hspace{2mm}$  $\forall$ $i \in \{1, 2, \dots n\}$.\\
where in feature space the data sample $x_i$ is represented by $\phi$ and the outlier is penalized by slack variable ${\xi}_i$. $\upsilon$ $\epsilon$ (0,1] decides the lower bound on the number of support vectors and upper bound on the fraction of outliers. 

The dual optimization problem of Eq.~\ref{eq2000} is defined as follows:

\begin{equation}
\centering
\min_{\alpha} \frac{1}{2} \sum\limits_{i=1}^{N}\sum\limits_{j=1}^{N}{\alpha}_i{\alpha}_jK(x_i,x_j)
\end{equation}
subject to: $0 \leq{\alpha}_i \leq \frac{1}{\upsilon N}$,    \   $\sum_{i=1}^{N}{\alpha}_i=1$, \   $i = \{1, 2, \dots, n\}$. \\
where $\alpha = {[ \alpha_1, \alpha_2,\dots \alpha_N]}^T$ and $\alpha_i$ is the Lagrange multiplier, whereas the weight-vector $w$ can be expressed as:
\begin{equation}
\centering
w=\sum\limits_{i=0}^{N}{\alpha_i \phi x(i)}
\end{equation}
$\rho$ is the margin parameter and computed by any $x_i$ whose corresponding Lagrange multiplier satisfies $0 < {\alpha}_i < \frac{1}{\upsilon N}$

\begin{equation}
\centering
\rho= \sum\limits_{j=1}^{N}{\alpha_j K(x_j, x_i)}
\end{equation}
The decision function with kernel expansion can be written as:

\begin{equation}
\centering
f(x) = \sum\limits_{i=1}^{N}{\alpha_i K(x_i, x)} - \rho
\end{equation}

Label of any test sample $x$ can be decided with the following:
\begin{equation}
\centering
\hat{y}= sign(f(x))
\end{equation}
where $sign(.)$ is sign function.

\begin{figure}
	\centering
	\includegraphics[width=\linewidth] {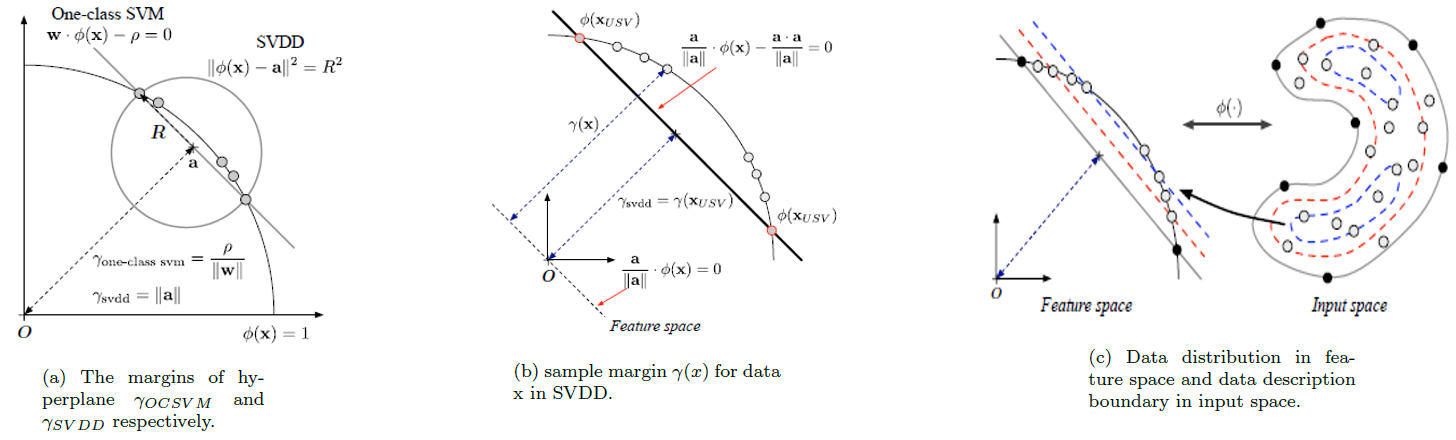}
	\caption{Representation of SVDD and OCSVM in unit norm space [35].}
	\label{fig42}
\end{figure}

It is observed that both SVDD and OCSVM perform same with Gaussian kernel and origin plays a key role because all outliers reside there. Norm of the centre of SVDD is equal to margin of a hyperplane of OCSVM in unit norm feature space is~\cite{kim2008fast} as depicted in Fig.~\ref{fig42}(a). Thus, reformulation of SVDD can be done by a hyperplane as follows:

\begin{equation}
\parallel \phi(x)-a\parallel^2\leq R^2 \Leftrightarrow w_{SVDD} . \phi(x)-\rho_{SVDD} \geq 0 
\end{equation}\\
where $a$ represents the the centre the hypersphere. $w_{svdd}$ and $\rho_{svdd}$ are the normal vector and the bias respectively of SVDD hyperplane and defined as below: 
\begin{equation}
w_{SVDD} =\frac{a}{\parallel a \parallel}, \hspace{5mm} \rho_{SVDD} =\parallel a \parallel
\end{equation}

In feature space, the virtual hyperplane passes through the origin and the sample margin is defined by its distance from the image of the data $\emph{x}$ as shown in Fig.~\ref{fig42}(b). The SVDD's sample margin is defined as below:
\begin{equation}
\gamma_{SVDD}(x)=\frac{a.\phi(x)}{\parallel a \parallel}
\end{equation}\\
where $\gamma$($\emph{x}$) is the image of data $\emph{x}$ in feature space and the sample margin of OCSVM is defined as follows:
\begin{equation}
\gamma_{OCSVM}(x)=\frac{w.\phi(x)}{\parallel w \parallel}
\end{equation}

Because data samples exist on the surface of a unit hypersphere, the maximum and minimum value of sample margin is defined as \enquote*{1} and \enquote*{0} respectively as follows:

\begin{equation}
0\leq \gamma(x) \leq 1
\end{equation}\\
Also, sample margin of unbounded support vectors  $x_{USV}$ ( 0 \textless $\alpha_{x_{USV}}$ \textless  $\frac{1}{\upsilon N}$) are the same as the margin of hyperplane, therefore:

\begin{equation}
\gamma(x_{USV})= \gamma_{SVDD}= \parallel a \parallel
\end{equation}
\begin{equation}
\gamma(x_{USV})= \gamma_{OCSVM}= \frac{\rho} {\parallel w \parallel}
\end{equation}

In feature space, the distribution of samples is represented by the sample margiin. Fig.~\ref{fig42}(c) shows the distribution of sample margin of training data and hyperplane of OCSVM in feature space. It is also observed that for RBF kernel both SVDD and OCSVM perform equally, hence in the present research, OCSVM is chosen for enhancement with the help of pinball loss function.

\subsubsection{OCSVM with other loss functions}

For every machine learning tasks such as classification, regression, etc., the loss functions plays vital role. Consider the training set $S = \{x_1, x_2, \dots x_n\}$ , where $x_i \in R^n$ are inputs and target class $y$ is defined as $y \in \{1\}$.
Let, $f : R^n \xrightarrow{} R$ is a mapping from $x_i \in R^n$ to $y \in \{1\}$. During training and testing the predicted output value of samples is supervised by the associated loss functions.

\begin{itemize}
 
\item \textit{OCSVM with hinge loss function:}

Like other machine learning tasks the hinge loss for OCSVM \cite{Xiaoramp} is defined as follows:

\begin{equation}
H_{loss} (z_i) = max ( 0 , \rho- z_i)
\end{equation}

This unbounded loss function is capable to deal with the shortest distance of the target class samples, but sensitive to the noise. The hinge loss OCSVM can be written as:

\begin{equation}
\label{eq222}
\max_{w,\xi , \rho}  \frac{1}{2} \parallel w \parallel^2 - \rho + \frac{1}{\upsilon N} \sum_{i}^{N} {H_{loss}(z_i)}
\end{equation}
where $z_i = w \phi(x_i)$. The constraints defined in Eq.~\ref{eq222} are integrated into $H_{loss}(z_i)$. The samples satisfying $z_i = w. \phi(x_i) \geq \rho$, are positioned above the hyperplane and no penalty is added, hence $H_{loss}(z_i) = 0$, whereas samples satisfying $z_i = w \phi(x_i) < \rho$ are
located on the other side of the hyperplane and some penalty should
be inflicted, therefore $H_{loss}(z_i) = \rho - z_i > 0$, which says that if the samples move far away from the hyperplane the penalty increases.

\item \textit{OCSVM with ramp loss function:}

As described above, the hinge loss function is unbounded and sensitive to the noise, the ramp loss function was introduced to OCSVM to make the model more robust \cite{Xiaoramp}. The ramp loss function is described as follows:

\begin{equation}
{R_{loss}}_{(\rho ,l )}(z_i) =
\begin{cases}
0, & z_i \geq \rho \\
\rho - z_i, & \rho l < z < \rho \\
\rho -l\rho, & z_i \leq \rho l
\end{cases}
\end{equation}
where $0 < l < 1$. When $z_i > l\rho$, the ramp loss is same to the hinge loss; whereas for $z_i \leq l\rho$, the ramp loss is
a constant that is an advantage over hinge loss which increases as $z_i$ decreases. This loss function is capable to deal with the shortest distance of the target class samples and less sensitive to the noise. The ramp loss OCSVM can written as:

\begin{equation}
\max_{w,\xi , \rho}  \frac{1}{2} \parallel w \parallel^2 - \rho + \frac{1}{\upsilon N} \sum_{i}^{N} {R_{loss}(z_i)}
\end{equation}
where $z_i = w \phi(x_i)$. Due to non-convexity of ramp loss the  OCSVM problem is no longer a convex optimization. This problem is solved by the phenomenon that the $R_{loss}$ is actually
the difference between two hinge loss functions and the modified formulation of the ramp loss OCSVM consists of the convex part and the concave part. This new problem is solved by \enquote{Concave-Convex
Procedure} (CCP) procedure \cite{collobert2006trading}.

\item \textit{Pin ball SVM:}
The pinball loss function is given as follows, which can be regarded as a generalized $l_1$ loss.

\begin{equation}
\centering
P_{\tau}(u) =
\begin{cases}
u, & u \geq 0 \\
 -\tau u, & u < 0
\end{cases}
\label{eq0001}
\end{equation}
where range of $\tau$ is [0, 1] \cite{huang2013support}. For quantile regression \cite{koenker2001quantile}, \cite{christmann2008svms}, the pinball loss $P_{\tau}$ has been successfully applied. 

Huang et al. \cite{huang2013support} derived quantile classification using the idea of support vector machine (SVM). It has been shown that pinball and hinge loss SVM have similar consistency property and computational complexity, moreover the pinball loss SVM is less sensitive to noise.

Huang et al. \cite{huang2013support} introduced pinball loss into the SVM. Consider the data points to be classified denoted by the set $T$. To solve the classification task, $w \in  R^n$ and $b \in R$ must be identified such that:
\begin{equation}
   \begin{aligned}
& w^T \phi (x_i) + b \geq 1 \hspace{.7cm} \text{for} \  y_i = 1 \\
& w^T \phi (x_i) + b \leq -1 \hspace{.5cm} \text{for} \ y_i = -1
\end{aligned}
\end{equation}
where $\phi(.)$ is a nonlinear mapping from the input space to a new feature space. The optimal hyperplane $w^T \phi(x) + b = 0$,
lies exactly between the supporting parallel hyperplanes given by:

\begin{equation}
\centering
w^T \phi(x) + b = 1 \ \text{and} \  w^T \phi(x) + b = -1
\label{eq7}
\end{equation}
and separates the binary class samples from each other with a margin of $\frac{1}{\parallel w \parallel}$ on each side. Data points residing on the supporting hyperplanes (Eq.~\ref{eq7}) are termed as support vectors. The classifier is obtained by maximizing the margin. If the data points of two different classes are not linearly separable in feature space then to separate all the data points correctly, pinball loss $P_{\tau} (x, y, f(x))$ is introduced to allow the existence of data points that violate the constraints $y_i(w^T \phi(x_i) + b) \geq 1$. Finally, the nonlinear Pin-SVM is formulated as follows:

\begin{equation}
\max_{w,b}  \frac{1}{2} \parallel w \parallel^2+ c \sum_{i=0}^{t} P_{\tau} (z_i)
\label{eq8}
\end{equation}

After employing the pinball loss function in Eq.~\ref{eq8}, following quadratic programming problem (QPP) come into existence for Pin-SVM:

\begin{equation}
\label{eq9}
\max_{w,\xi , \rho}  \frac{1}{2} \parallel w \parallel^2+ c \sum_{i=1}^{t} \xi_i 
\end{equation}
s.t. 
\begin{equation}
\label{eq10}
\begin{aligned}
& y_i(w^T \phi(x_i) + b) \geq 1 - \xi_i,\\
& y_i(w^T \phi(x_i) + b) \leq 1 + \frac{\xi_i}{\tau} , i = 1, 2, \dots N
\end{aligned}
\end{equation}
where $\xi = (\xi_1, \xi_2, . . ., \xi_N )^T$ is a slack variable and $c >$ 0 is a penalty parameter. The parameter c determines the weight between the two terms $\parallel w \parallel^2$ and
$\sum_{i=1}^{t} \xi_i$. Pin-SVM has an advantage of noise insensitivity. Meanwhile, Pin-SVM has similar time complexity to that of standard SVM. Note that the second constraint of Eq.~\ref{eq10} becomes $\xi \geq 0$ when $\tau = 0$, and thus Pin-SVM reduces to the hinge loss SVM.

\end{itemize}

\subsection{OCSVM with pinball loss function (PB-OCSVM)}

From above discussion it is clear that the hinge loss is sensitive to the noise. To overcome from this problem
and to improve the performance of OCSVM, pinball loss function \cite{huang2013support} is fused to standard
OCSVM to obtain pinball OCSVM (PB-OCSVM). This enhanced version deals with quantile distance \cite{koenker2001quantile} that reduces the sensitivity to the noise. After introducing pinball loss function, the conventional OCSVM can be rewritten as follows.

\begin{equation}
\label{eq00002}
\max_{w,\xi , \rho}  \frac{1}{2} \parallel w \parallel^2 - \rho + \frac{1}{\upsilon N} \sum_{i}^{N} P_{\tau}(z_i) 
\end{equation}
where $P_{\tau}$ is pinball loss as shown in Eq.~\ref{eq0001} with $z_i= w^T \phi(x_i)$ which helps to the slope on the target class. Substituting the pinball loss into Eq.~\ref{eq00002}, following QPP is obtained:

\begin{equation}
\label{eq000002}
\begin{aligned}
  \max_{w,\xi , \rho}  \frac{1}{2} \parallel w \parallel^2 - \rho + \frac{1}{\upsilon N} \sum_{i}^{N} {\xi}_i \\ 
\text{subject to: \hspace{1cm}}  w.\phi(x_i) \geq \rho - {\xi}_i \\
\hspace{2cm} w.\phi(x_i) < \rho +\frac{{\xi}_i}{\tau}
\end{aligned}
\end{equation}

\subsubsection{Dual problem and kernel formulation}

Now after introducing a kernel based formulation to the
PB-OCSVM, the Lagrangian with ${\alpha}_i \geq 0$, ${\beta}_i \geq 0$ of Eq.~\ref{eq000002} is as follows:

\begin{equation}
    \centering
    \begin{aligned}
    \mathcal{L} (w, b, \xi , \alpha, \beta)
= &\frac{1}{2} {\parallel w \parallel}^2 + c \sum_{i=1}^{N} {\xi}_i - \sum_{i=1}^{N} {\alpha}_i (w.\phi(x_i)- \\ & \rho+ {\xi}_i)- \sum_{i=1}^{N} {\beta}_i (w.\phi(x_i)-\rho- \frac{{\xi}_i}{\tau}) 
\end{aligned}
    \label{eq003}
\end{equation}
According to:\\
$\frac{\delta \mathcal{L}}{\delta w} = w + \sum_{i=1}^{N} {\alpha}_i \phi(x_i) =0$, \\
$\frac{\delta \mathcal{L}}{\delta \rho} =  \sum_{i=1}^{N} {\alpha}_i -{\beta}_i  =0$, \\
$\frac{\delta \mathcal{L}}{\delta \xi} =c - {\alpha}_i - \frac{1}{\tau} {\beta}_i =0, \forall i= 1,2, \dots, N $

the dual problem of Eq.~\ref{eq003} is obtained as follows:
\begin{equation}
\centering
\begin{aligned}
&\max_{\alpha, \beta} - \frac{1}{2} \sum\limits_{i=1}^{N}\sum\limits_{j=1}^{N}({\alpha}_i-{\beta}_i) \phi(x_i)^T \phi (x_j) ({\alpha}_j-{\beta}_j) +\sum\limits_{i=1}^{N}({\alpha}_i-{\beta}_i) \\
&\text{subject to:} \\
&\sum\limits_{i=1}^{N}({\alpha}_i-{\beta}_i)=0,\\
&{\alpha}_i+ \frac{1}{\tau}{\beta}_i =c, i= 1, 2, \dots N,\\
&{\alpha}_i \geq 0, {\beta}_i \geq 0, i = 1, 2, \dots N
\end{aligned}
\label{eq0004}
\end{equation}

Introducing the positive definite kernel $K(x_i, x_j) =
\phi(x_i)^T \phi (x_j)$ and variables $\lambda_i = {\alpha}_i-{\beta}_i$, we get

\begin{equation}
\centering
\begin{aligned}
&\max_{\alpha, \beta} - \frac{1}{2} \sum\limits_{i=1}^{N}\sum\limits_{j=1}^{N}\lambda_i K(x_i, x_j) \lambda_j +\sum\limits_{i=1}^{N}\lambda_i \\
&\text{subject to:} \\
&\sum\limits_{i=1}^{N} \lambda_i=0,\\
&{\lambda}_i+ (1+\frac{1}{\tau}){\beta}_i =c, i= 1, 2, \dots N,\\
&{\lambda}_i+ \beta_i \geq 0, {\beta}_i \geq 0, i = 1, 2, \dots N
\end{aligned}
\label{eq0005}
\end{equation}

With kernel expansion the decision function can be defined as follows:

\begin{equation}
\centering
\mathcal{f}(x) = \sum\limits_{i=1}^{N}{\lambda_i K(x_i, x)} - \rho
\end{equation}
Finally, the test instance $x$ can be labelled as follows:
\begin{equation}
\centering
\hat{y}= sign(\mathcal{f}(x))
\end{equation}
where $sign(.)$ is sign function.

It is well known that the computational complexity of OCSVM is $\mathcal{O}(N_{T_c}^3)$ where $N_{T_c}$ is the number of the only target class samples, whereas the computational complexity of PB-OCSVM is found same as OCSVM i.e. $\mathcal{O}({N_{T_c}}^3)$. Therefore, the complexity of the PB-OCSVM is equal to conventional OCSVM, hence pinball loss function does not increase the computation time of PB-OCSVM.

\section{Experiments and results}
\label{l5}

COVID-19 positive samples of COVID-19 dataset \cite{cohen2020covid} are used as training samples for OCSVM and the proposed PB-OCSVM whereas other samples of this dataset along with samples of datasets RSNA \cite{Stein} and U.S. national library of
medicine (USNLM) collected Montgomery country - NLM(MC) \cite{jaeger2014two} are used for testing. All the images of fused dataset is pre-processed as discussed in Subsection \ref{pre} that outputs the images of size $331 \times 331 \times 3$, where $3$ is the number of channels. For PB-OCSVM and OCSVM, the COVID-19 positive samples are considered as training samples whereas deep learning models are trained with multi-class samples (80\% from each class) obtained via respective class balancing techniques. For OCC approaches, all in-class and other class samples are treated as test samples and for deep learning models all training, remaining in-class and other class samples are treated as test samples. Once training is done with suitable parameters, OCSVM and PB-OCSVM are used for testing in two scenarios: test with training samples and other samples of all three datasets. For performance evaluation five benchmark parameters have been used to show effectiveness of the proposed model: accuracy, precision, sensitivity, specificity and confidential interval (CI) with 95\% and 98\% confidence (Eq.~\ref{eq56}). Comparison of performance of OCSVM and PB-OCSVM is done with recently proposed deep learning models as shown in Table~\ref{tab111}. It is known that all deep learning models need huge amounts of data and at the same time all classes must be well defined. Hence, it is necessary to oversampling of COVID-19 samples to any deep learning model. It is observed that all recent deep learning architectures \cite{oh2020deep}, \cite{afshar2020covid}, \cite{wang2020covid}, \cite{apostolopoulos2020covid}, \cite{hall2020finding} used oversampling to deal with the data imbalance problem and after massive computational efforts were able to give significant performance.

\begin{equation}
\centering
\begin{aligned}
&Accuracy =  \frac{TN+TP}{TN+FN+TP+FP} \\
&Precision =\frac{TP}{TP+FP} \\
&Specificity= \frac{TN}{TN+FP} \\
&Sensitivity=\frac{TP}{TP+FN}\\
&Confidence\_Interval= z *(\sqrt{accuracy* (1-accuracy)/N}\\
\end{aligned}
\label{eq56}
 \end{equation}
 where  $z$ is standard deviation (1.96 for 95\% confidence and 2.33 for 98\% confidence) and $N= 1214$ is number of samples.

There are total four classes in fused dataset: COVID-19, normal, Tuberculosis and other pneumonia. Some existing models have been proposed considering all four classes whereas some articles considered Tuberculosis and other pneumonia classes as a single class that gives three classes: COVID-19, normal and other pneumonia. Few papers considered normal, Tuberculosis and other pneumonia as non COVID class that gives two classes COVID and non COVID. For existing deep learning architectures, experiments have been performed in three categories with consideration of 4, 3 and 2 classes. For exhaustive comparative analysis, all scenarios have also been considered for OCSVM and PB-OCSVM and it is observed that both the models exhibited the consistent performance due of their working principle. Whereas it is identified that deep learning models perform better in case of binary class classification approach. It is also observed that both adaptive total variation and ACE performed nearly equal.

\begin{table}[]
\centering
\caption{Performance comparison.}
\resizebox{\textwidth}{!}{%
\begin{tabular}{|c|c|c|c|c|c|c|c|c|c|c|c|c|}
\hline
\multirow{2}{*}{\textbf{Authors}}                                     & \multirow{2}{*}{\textbf{Technique}} & \multirow{2}{*}{\textbf{Classes}} & \multicolumn{5}{c|}{\textbf{Adaptive total variation}}                                  & \multicolumn{5}{c|}{\textbf{Automatic color equalization}}                              \\ \cline{4-13} 
                                                                      &                                     &                                   & A             & P             & S             & Sp            & 95\% \& 98\% CI         & A             & P             & S             & Sp            & 95\% \& 98\% CI         \\ \hline
Yujin et al. \cite{oh2020deep}                       & Fine Tuning                         & \multirow{4}{*}{4}                & 0.88          & 0.83          & 0.86          & 0.96          & $\pm$ (1.83\% - 2.17\%) & 0.86          & 0.86          & 0.88          & 0.92          & $\pm$ (1.95\% - 2.32\%) \\ \cline{1-2} \cline{4-13} 
Afshar et al. \cite{afshar2020covid}                 & Fine Tuning                         &                                   & 0.95          & 0.92          & 0.90          & 0.95          & $\pm$ (1.10\% - 1.31\%) & 0.94          & 0.94          & 0.92          & 0.91          & $\pm$ (1.36\% - 1.58\%) \\ \cline{1-2} \cline{4-13} 
OCSVM                                                                 &                                     &                                   & 0.97          & 0.97          & 0.93          & 0.96          & $\pm$ (0.96\% - 1.14\%) & 0.96          & 0.96          & 0.94          & 0.95          & $\pm$ (1.10\% - 1.31\%) \\ \cline{1-2} \cline{4-13} 
PB-OCSVM                                                              &                                     &                                   & \textbf{0.98} & \textbf{0.99} & \textbf{0.98} & \textbf{0.98} & $\pm$ (0.79\% - 0.94\%) & \textbf{0.97} & \textbf{0.98} & \textbf{0.98} & \textbf{0.98} & $\pm$ (0.95\% - 1.14\%) \\ \hline
Wang et al. \cite{wang2020covid}                     & Fine Tuning                         & \multirow{4}{*}{3}                & 0.92          & 0.91          & 0.88          & 0.89          & $\pm$ (1.53\% - 1.81\%) & 0.92          & 0.91          & 0.9           & 0.91          & $\pm$ (1.53\% - 1.81\%) \\ \cline{1-2} \cline{4-13} 
Apostolopoulos et al. \cite{apostolopoulos2020covid} & Fine Tuning                         &                                   & 0.87          & 0.93          & 0.92          & \textbf{0.98} & $\pm$ (1.89\% - 2.25\%) & 0.89          & 0.94          & 0.93          & 0.95          & $\pm$ (1.76\% - 2.09\%) \\ \cline{1-2} \cline{4-13} 
OCSVM                                                                 &                                     &                                   & 0.97          & 0.97          & 0.93          & 0.96          & $\pm$ (0.96\% - 1.14\%) & \textbf{0.96} & \textbf{0.96} & 0.94          & 0.96          & $\pm$ (1.10\% - 1.31\%) \\ \cline{1-2} \cline{4-13} 
PB-OCSVM                                                              &                                     &                                   & \textbf{0.98} & \textbf{0.99} & \textbf{0.98} & \textbf{0.98} & $\pm$ (0.79\% - 0.94\%) & \textbf{0.96} & \textbf{0.96} & \textbf{0.95} & \textbf{0.97} & $\pm$ (1.10\% - 1.31\%) \\ \hline
Hall et al. \cite{hall2020finding}                   & Fine Tuning                         & \multirow{4}{*}{2}                & 0.91          & 0.88          & 0.88          & 0.93          & $\pm$ (1.61\% - 1.91\%) & 0.89          & 0.89          & 0.88          & 0.93          & $\pm$ (1.76\% - 2.09\%) \\ \cline{1-2} \cline{4-13} 
Apostolopoulos et al. \cite{apostolopoulos2020covid} & Fine Tuning                         &                                   & \textbf{0.98} & 0.97          & 0.92          & \textbf{0.98} & $\pm$ (0.79\% - 0.94\%) & 0.96          & 0.95          & 0.92          & 0.95          & $\pm$ (1.10\% - 1.31\%) \\ \cline{1-2} \cline{4-13} 
OCSVM                                                                 &                                     &                                   & 0.97          & 0.97          & 0.93          & 0.96          & $\pm$ (0.96\% - 1.14\%) & 0.95          & 0.95          & 0.94          & 0.94          & $\pm$ (1.22\% - 1.46\%) \\ \cline{1-2} \cline{4-13} 
PB-OCSVM                                                              &                                     &                                   & \textbf{0.98} & \textbf{0.99} & \textbf{0.98} & \textbf{0.98} & $\pm$ (0.79\% - 0.94\%) & \textbf{0.97} & \textbf{0.98} & \textbf{0.98} & \textbf{0.98} & $\pm$ (0.95\% - 1.14\%) \\ \hline
\multicolumn{13}{l}{* A- Accuracy, P- Precision, S- Sensitivity, Sp- Specificity}
\end{tabular}%
}
\label{tab111}
\end{table}

\subsection{Workability of proposed PB-OCSVM in presence of noise}

To validate the robustness of the proposed PB-OCSVM against the noise, and to demonstrate its generalized behaviour experiments are performed considering following scenarios:
\begin{itemize}
    \item [\textit{a}.] Images obtained after inpainting operation as discussed in Subsection \ref{pre}.
    \item [\textit{b}.] Images generated after introducing  Gaussian, laplacian and uniform noise to the final images obtained by adaptive total variation method. 
    \item [\textit{c}.] Experiments with 10 benchmark UCI datasets.
    \end{itemize}
    
For scenarios $a$ and $b$ all state-of-the-art methods are evaluated along with proposed method whereas for scenario $c$ conventional OCSVM and proposed PB-OCSVM are evaluated. In scenario $a$ the images obtained after binary thresholding as discussed in Subsection \ref{pre} are used for computation. The partially processed images are used by all the above discussed classifiers. Whereas, to illustrate the effectiveness of the proposed method, the output images obtained after pre-processing are further fused with Gaussian, laplacian and uniform noise are used in scenario $b$. Table \ref{tab3} shows the comparative performance of state-of-the-art approaches and proposed approach (best results are represented in bold). To tune the hyperparameters of OCSVM and PB-OCSVM, grid search mechanism is used. Results show that the in the presence of noise, the proposed method outperformed compared to deep learning models and conventional OCSVM.

\begin{table}[H]
\centering
\caption{Performance comparison in noisy environment}
\resizebox{\linewidth}{!}{
\begin{tabular}{cc|c|c|c|c|c|c|c|}
\cline{3-9}
\textbf{} &
  \textbf{} &
  \multicolumn{4}{c|}{\multirow{2}{*}{\textbf{\begin{tabular}[c]{@{}c@{}}Scenario-a (Partially \\ processed images)\end{tabular}}}} &
  \multicolumn{3}{c|}{\textbf{Scenario-b (in presence of noise)}} \\ \cline{1-2} \cline{7-9} 
\multicolumn{1}{|c|}{\multirow{2}{*}{\textbf{Authors}}} &
  \multirow{2}{*}{\textbf{Classes}} &
  \multicolumn{4}{c|}{} &
  \multicolumn{3}{c|}{\textbf{AUC}} \\ \cline{3-9} 
\multicolumn{1}{|c|}{} &
   &
  \textbf{A} &
  \textbf{P} &
  \textbf{S} &
  \textbf{Sp} &
  \textbf{Gaussian} &
  \textbf{Laplacian} &
  \textbf{Uniform} \\ \hline
\multicolumn{1}{|c|}{Yujin et al. \cite{oh2020deep}}          & \multirow{4}{*}{4} & 0.82                           & \textbf{0.86}                           & 0.87                           & \textbf{0.89}                           & 0.83                & 0.84                 & 0.86               \\ \cline{1-1} \cline{3-9} 
\multicolumn{1}{|c|}{Afshar et al. \cite{afshar2020covid} }         &                    & 0.83                           & 0.81                           & 0.86                           & 0.82                           & 0.79                & 0.78                 & 0.81               \\ \cline{1-1} \cline{3-9} 
\multicolumn{1}{|c|}{OCSVM}                 &                    & 0.82                           & 0.81                           & 0.85                           & 0.83                           & 0.84                & 0.83                 & \textbf{0.88}               \\ \cline{1-1} \cline{3-9} 
\multicolumn{1}{|c|}{PB-OCSVM}              &                    & \textbf{0.85}                           & \textbf{0.86}                           & \textbf{0.88}                          & \textbf{0.89}                          & \textbf{0.88}                & \textbf{0.89}                 & \textbf{0.88}               \\ \hline
\multicolumn{1}{|c|}{Wang et al. \cite{wang2020covid}}           & \multirow{4}{*}{3} & 0.82                           & 0.82                           & 0.85                           & 0.84                           & 0.84                & 0.85                 & 0.84               \\ \cline{1-1} \cline{3-9} 
\multicolumn{1}{|c|}{Apostolopoulos et al. \cite{apostolopoulos2020covid}} &                    & 0.85                           & 0.84                           & 0.85                           & 0.81                           & 0.75                & 0.79                 & 0.81               \\ \cline{1-1} \cline{3-9} 
\multicolumn{1}{|c|}{OCSVM}                 &                    & 0.82                           & 0.81                           & 0.85                           & 0.83                           & 0.83                & \textbf{0.89}                 & 0.84               \\ \cline{1-1} \cline{3-9} 
\multicolumn{1}{|c|}{PB-OCSVM}              &                    & \textbf{0.88}                           & \textbf{0.86}                           & \textbf{0.89}                           & \textbf{0.91}                           & \textbf{0.89}                & \textbf{0.89}                & \textbf{0.88}               \\ \hline
\multicolumn{1}{|c|}{Hall et al. \cite{hall2020finding} }           & \multirow{4}{*}{2} & 0.81                           & 0.83                           & 0.82                           & 0.84                           & 0.81                & 0.82                 & 0.84               \\ \cline{1-1} \cline{3-9} 
\multicolumn{1}{|c|}{Apostolopoulos et al. \cite{apostolopoulos2020covid}} &                    & 0.79                           & 0.82                           & 0.84                           & 0.85                           & 0.76                & 0.78                 & 0.81               \\ \cline{1-1} \cline{3-9} 
\multicolumn{1}{|c|}{OCSVM}                 &                    & 0.86                           & 0.86                           & 0.85                           & 0.87                           & 0.83                & 0.86                 & 0.84               \\ \cline{1-1} \cline{3-9} 
\multicolumn{1}{|c|}{PB-OCSVM}              &                    & \textbf{0.91}                           & \textbf{0.89}                          & \textbf{0.91}                           & \textbf{0.92}                           & \textbf{0.88}                & \textbf{0.89}                &  \textbf{0.89}             \\ \hline
\multicolumn{9}{l}{* A- Accuracy, P- Precision, S- Sensitivity, Sp- Specificity}
\end{tabular}}
\label{tab3}
\end{table}

To illustrate the generalized effectiveness of the proposed model, 10 benchmark UCI datasets are experimented with OCSVM and PB-OCSVM. Table \ref{tab4} shows the description of datasets used and AUC score of both the classifiers. Experimental results show that the PB-OCSVM performs marginally better than the conventional OCSVM. To ensure the workability in minimum number of samples, 80\% of the target class samples are randomly chosen for training whereas all target class samples and other class instances are selected for testing as done in earlier experiments.

\begin{table}[H]
\centering
\caption{Experimental results with UCI datasets}
\resizebox{\textwidth}{!}{%
\begin{tabular}{|c|c|c|c|c|c|c|c|}
\hline
\multicolumn{1}{|c|}{\multirow{2}{*}{\textbf{Dataset}}} & \multirow{2}{*}{\textbf{N\textsubscript{Target}}} & \multirow{2}{*}{\textbf{N\textsubscript{Outliers}}} & \multirow{2}{*}{\textbf{Attributes}} & \multirow{2}{*}{\textbf{N\textsubscript{Training}}} & \multirow{2}{*}{\textbf{N\textsubscript{Test}}} & \multicolumn{2}{c|}{\textbf{AUC}}  \\ \cline{7-8} 
\multicolumn{1}{|c|}{}                                  &                                     &                                       &                                      &                                       &                                   & \textbf{OCSVM} & \textbf{PB-OCSVM} \\ \hline
Blood Transfusion                                       & 178                                 & 570                                   & 4                                    & 142                                   & 606                               & \textbf{0.86}           & 0.85              \\ \hline
Wholesale Customers                                     & 298                                 & 142                                   & 7                                    & 238                                   & 202                               & 0.94           & \textbf{0.96}              \\ \hline
Breast Cancer                                           & 77                                  & 186                                   & 9                                    & 62                                    & 201                               & 0.93           & \textbf{0.96}              \\ \hline
Glass                                                   & 70                                  & 77                                    & 10                                   & 56                                    & 91                                & \textbf{0.98}           & \textbf{0.98}              \\ \hline
Heart                                                   & 120                                 & 150                                   & 13                                   & 96                                    & 174                               & 0.95           & \textbf{0.96}              \\ \hline
Climate Model                                           & 294                                 & 46                                    & 18                                   & 235                                   & 105                               & 0.94           & \textbf{0.98}              \\ \hline
Hepatitis                                               & 123                                 & 32                                    & 19                                   & 98                                    & 57                                & \textbf{0.95}           & \textbf{0.95}              \\ \hline
Parkinsons                                              & 147                                 & 48                                    & 22                                   & 118                                   & 77                                & 0.97           & \textbf{0.98}              \\ \hline
QSAR biodegradation                                     & 356                                 & 699                                   & 41                                   & 285                                   & 770                               & \textbf{0.95}           & 0.94              \\ \hline
Sonar                                                   & 111                                 & 97                                    & 60                                   & 89                                    & 119                               & 0.93           & \textbf{0.95}              \\ \hline
\multicolumn{8}{l}{* N\textsubscript{Target}: number of target samples, N\textsubscript{Outliers}: number of other class samples etc.}
\end{tabular}}
\label{tab4}
\end{table}

\section{Conclusion}
\label{l6}

The present research proposes a novel pinball OCSVM (PB-OCSVM) for early-detection of COVID-19 in presence of limited number of samples. Recently, several deep learning based automated approaches have been proposed for diagnosis of COVID-19 using CXR images. All these solutions need enormous multi-class data samples for unbiased operation. Due to limited availability of CXR images, it is evident that all deep learning approaches use collection of multiple datasets and prefer oversampling strategies to ensure smooth functioning. The COVID-19 is a newly known deadly pandemic and its precise characteristics are still unknown; therefore, oversampling must not be considered as a concrete solution to achieve higher diagnosis accuracy. 

In present research, the COVID-19 diagnosis is considered as an anomaly/ novelty detection task and a novel variant of OCSVM is proposed under the supervision of pinball loss function called PB-OCSVM capable to work with limited number of COVID-19 infected CRX images. The proposed model does not add extra computation overhead and ensures quicker diagnosis. For experiments collection of multiple datasets are used where the available COVID-19 infected X-ray images are used for training whereas all in-class and other class samples are treated as test samples. The same fused dataset is used to evaluate state-of-the-art deep learning models and conventional OCSVM. Experimental results show that PB-OCSVM outperformed state-of-the-art deep learning models and conventional OCSVM. Meanwhile, it is also evident that the proposed model ensures negligible false-positive and false-negative rates compared to OCSVM and attains significantly better accuracy, precision, specificity and sensitivity. 

To validate the performance and robustness of PB-OCSVM, experiments are performed with noisy images. Two noisy scenarios are considered for experiments and results show that PB-OCSVM performs better in both the scenarios compared to state-of-the-art approaches. For generalized performance evaluation of proposed model, experiments are also performed with 10 benchmark UCI datasets. Experimental results proves that PB-OCSVM performs significantly better than conventional OCSVM.

The disease diagnosis in healthcare domain is especially an anomaly or novelty detection task where one-class classification approaches are proven more promising compared to conventional binary or multi-class classification approaches. In this context, it is believed that the application of proposed model is not limited and can also be extended to other disease diagnosis tasks. Meanwhile, the same work can also be extended to other application domains for anomaly/novelty detection objectives.

\bibliographystyle{ws-ijprai}
\bibliography{mybibfile}

\end{document}